# SPECIAL RELATIVITY IS CONSISTENT WITH THE OPERA MEASUREMENTS OF THE NEUTRINO "VELOCITY"

Jean Paul Mbelek[1]

1. Sangha Center for Astronomy, Astrophysics and Cosmology, Sangha, Mali

Abstract : We show that special relativity (SR) may be consistent with the OPERA measurements of the neutrino velocity provided the latter is corrected for the second order term in $V^2/c^2$ implied by the velocity, V, of the α particles from radioactive rocks of the experiment area, when properly accounted for in the SR velocity addition law. An upper bound has been set on the velocity of the OPERA neutrinos by using the deformed dispersion relation suggested by the result of the experiment OPERA itself.

I. Introduction

The OPERA collaboration recently claimed [1] that their precise measurement of the time of flight (TOF) of neutrinos and the distance from the CERN to the Gran Sasso imply for the speed, v, of muon neutrinos the estimate

$(v - c)/c = (2.48 ± 0.28 \text{ (stat)} ± 0.30 \text{ (sys)}) × 10^{-5}$.     (1)

The above result is derived subsequent to the measurement at 6σ level of an early arrival time, δt = (60.7 ± 6.9 (stat) ± 7.4 (sys)) ns, of muon neutrinos with respect to the one computed assuming the speed of light in vacuum, c.

Clearly, the OPERA results are problematic since superluminal neutrinos would violate Einstein causality. Now, as one knows, experiments designed to test Bell inequalities [2] have all strengthened quantum mechanics with great precision so far. Indeed, all these experiments favoured Einstein causality, non-locality and quantum non-separability at the expense of theories involving hidden variables [3].

In addition, it is worth noticing that superluminal neutrinos would also make fundamental interactions be superluminal by allowing non-causal interactions in contradiction with all known well-established experimental facts confirming general relativity (GR) and quantum field theory, namely QED, QCD and the electroweak theory (EW) all based on SR. Thus, for instance, any $W^±$ boson could decay into a charged lepton and a superluminal neutrino, thereby involving a non-causal interaction term in the EW Lagrangian density. Also, since photons still move with the speed of light, a photon could decay into a pair of a superluminal neutrino and an infraluminal antineutrino, then the superluminal neutrino would involve a non-causal interaction term in the QED Lagrangian density too, which contradicts the stringent upper bounds on the photon mass [4]. Generally speaking, within the present day paradigm of quantum field theory, tachyons (superluminal particles) are avoided because they give rise to violation of causality and unitarity, but see [5].

Moreover, the relative precision of the OPERA result is only 25 ppm whereas GR and QED which both exclude tachyons are validated with a relative precision which may reach $10^{-5}$ ppm. So, if founded, the OPERA claim still needs to be confirmed with a very high precision too. Indeed, because of all the numerous and impressive precision tests that have so far validated Einstein theory of relativity, it is not so simple to claim its break down. Instead, that goes without saying that it is generally preferred to invoke dark matter (DM) and dark energy (DE) at the expense of all other alternative theories like the modified Newtonian dynamics (MOND) or tensor-vector-scalar gravity (TeVeS) [6]. Also, it is odd to think that the Global Positioning System (GPS) which must take into account effects predicted by Einstein theory of relativity (both SR and GR) to achieve the desired 20-30 nanosecond accuracy is used to challenge one of the basic principles of this fundamental theory.

Some attempts to explain the OPERA results invoke large extra-dimensions and shortcuts through the bulk [7]. However, let us recall that unless a tunnelling effect is proved to be effective, in which case it should have shown up with the SN 1987A antineutrinos despite their lower energy (E = 10 MeV) but because of their longer TOF, the energy gap between the 4D-brane and the bulk is at least equal to 1 TeV if not the Planck energy which in both case is well beyond the energy of the neutrinos of the OPERA experiment. Moreover, the Heisenberg uncertainty relation $\delta E \times \delta t \approx \hbar$ would yield, for a transit time in the bulk $\delta t = 61$ ns, an energy fluctuation $\delta E \approx 1.1 \times 10^{-8}$ eV which is by far too weak for the OPERA experiment neutrinos travel through the bulk by quantum fluctuations.

Let us emphasize that a close look at the OPERA results not only might suggest superluminal neutrinos but also reveals the deformed dispersion relation that could be associated to them within the energy range of the experiment, namely some tens of GeV. Indeed, no clues on a possible energy dependence of $\delta t$ in the energy range explored by OPERA, within the statistical accuracy of the measurement was found. So, as a first approximation, combining the special relativistic relation $v = Pc^2/E$ with relation (1) yields the deformed dispersion relation

$$E^2 - P^2 c^2 = -\tfrac{1}{2}\,\delta\, E\,(E + P\,c), \qquad (2)$$

where $\delta \approx 5 \times 10^{-5}$ is a non-dimensional constant in the energy range of interest.

By comparing relation (2) with relation (11) of ref. [8] and assuming a left-handed neutrino (negative helicity), since $m_\nu \ll E/c^2$, it follows,

$$\eta_+ = 0 \text{ and } \delta = \eta_-\, E/E_{Planck}, \qquad (3)$$

where $E_{Planck} = 10^{19}$ GeV is the Planck energy scale.

According to Jacobson et al. (ref. [8], § 5.6), the parameter $\eta_-$ is constrained to $|\eta_-| \leq 0.2$ by fermion pair emission at 50 TeV (assuming stable Dirac neutrinos), hence relation (3) implies $\delta \ll 10^{-15}$. Clearly, the constraint on $\delta$ should be even stronger at lower energy as in the case of the OPERA experiment neutrinos (E = 13.9 GeV - 42.9 GeV). The latter independent constraint stand in support of the claim by Cohen and Glashow who put forward $\delta \ll 1.7 \times 10^{-11}$ and refute the superluminal interpretation of the OPERA result [9].

Hence we are faced with the recurrent question in metrology, namely what is exactly measured primarily and how this measurement is achieved in a given experiment. The answer is not always obvious but it should be found before one goes further to the interpretation of the experimental results. It turns out that both the putative time of flight, t, of the neutrinos and most of the distance, d, from the benchmarks at CERN to the OPERA detector at LNGS are derived from time measurements based on clocks synchronization with the help of the GPS clocks. This might be the weak spot of the claim for superluminal neutrinos by the OPERA collaboration. In what follows, we show that the influence of the α particles from radioactive rocks of the experiment area on the GPS communication signals may not have been fully taken into account. As a consequence, it follows in the clocks synchronized to the GPS, a systematic delay when compared to any other signal which is protected from the influence of extraneous particles. Hereafter, we explore the possibility that a systematic effect might enter in the OPERA measurement that is independent not only of the energy but of the neutrino behavior too.

II. The kinematic effect in excess on the communication signals of high Earth orbit satellites

In a previous work, we addressed the flyby anomaly [10] and show that it could be solved by the SR transverse Doppler effect related to the speed of Earth's rotation [11]. It may seem strange that such a well known SR effect be unaccounted for in the spacecraft communication signals. However, as one knows any relative motion has an impact on the frequency of a photon according to the accuracy of its measurement. To start with, let us consider a set of relative motions labelled (i) and each associated to a velocity $\mathbf{V}_i$ with respect to the same given reference frame. Considering the photons of the GPS communication signals, since they propagate through the Earth's atmosphere and the Earth's magnetosphere, a correction due to the α particles from radioactive rocks of the experiment area, by analogy with the Fizeau water experiment, should be made according to the SR theorem of the addition of velocities.

Let $\mathbf{V}_\gamma$ be the uplink or downlink velocity of the communication signals between the two experiment sites and the GPS satellites that are used to perform the synchronization of the CERN and NLGS clocks dedicated to the OPERA measurements of the neutrino velocity,

$$\mathbf{V}_\gamma = ((c/n)\,\mathbf{u} + \Sigma_i \mathbf{V}_i)/(1 + (c/n)\,\mathbf{u}.\Sigma_i \mathbf{V}_i/c^2), \quad (4)$$

where n denotes the refractive index of the atmosphere or the magnetosphere (n – 1 = 2.926 × $10^{-4}$ in the normal atmospheric conditions and less in the magnetosphere) and $\mathbf{u}$ is the unit vector in the direction of motion of the photons of the communication signal. Relation (4) may rewrite

$$\mathbf{V}_\gamma = (\mathbf{V}_{\gamma/0} + \mathbf{V})/(1 + (c/n)\,\mathbf{u}.\mathbf{V}/c^2), \quad (5)$$

where $\mathbf{V} = \mathbf{V}_0/\Gamma$, $\Gamma = 1 + \Sigma_{i\neq 0}\,\mathbf{u}.\mathbf{V}_i/nc$ and $\mathbf{V}_\gamma$ reduces to $\mathbf{V}_{\gamma/0}$ when the contribution of the velocity of the α particles is neglected,

$$\mathbf{V}_{\gamma/0} = ((c/n)\,\mathbf{u} + \Sigma_{i\neq 0}\,\mathbf{V}_i)/(1 + (c/n)\,\mathbf{u}.\Sigma_{i\neq 0}\,\mathbf{V}_i/c^2). \quad (6)$$

Since $V_{\gamma/0} \approx c/n$, one finds,

$$V_\gamma \approx V_{\gamma/1} - (c/n) \times (1 - (1/n^2)) \times (V^2/c^2), \quad (7)$$

where we have set $V_{\gamma/1} = (c/n) + V\,(1 - (1/n^2))$. As one can notice, usually the second order term $V^2/c^2$ is neglected in relation (7) by assuming $V \ll c$. This approximation is even more justified in the Earth's atmosphere or the interplanetary medium, because of the suppression factor $1 - (1/n^2)$. For this reason, we consider that $V_{\gamma/1}$ stands in a manner for the velocity of the photons as computed by the GPS system, so that the effect of the α particles is taken into account in the first order approximation of V/c by the GPS. Now, the GPS satellites are orbiting at the altitude h = 20 200 km in the Earth's magnetosphere which extends outward the Earth between 450 km and about 60 000 km from the surface of the Earth. So, the communication signals travel a distance $D \approx h$ ($D = (h^2 + x^2)^{1/2}$, with $0 \leq x \leq d$) through the flow of α particles during their travel between the experiment sites and the GPS satellites. Consequently, in as much as the second order term depending on the velocity $V_0$ would not be included to correct the velocity of the communication signals, a systematic error, δt, in the time measurements would be involved as a delay in excess by referring to the GPS clocks. Hence, since the same error does not affect the velocity, v, of the neutrinos travelling a distance, d, deep enough through the Earth's crust, one would be inclined to conclude that neutrinos are travelling faster than light with a relative difference in velocity $(v - c)/c = δt/t$, where $t = d/v$.

Thus, by taking thoroughly into account the correction to the TOF of the photons due to the flow of α particles, the one-way delay in excess of the communication signal between the experiment sites and the GPS satellites reads

$$δt = (D/V_\gamma) - (D/V_{\gamma/1}) \approx n \times (1 - (1/n^2)) \times DV^2/c^3 \approx 2(n-1) \times hV_0^2/c^3. \quad (8)$$

One finds in text books the average velocity of α particles from radioactive rocks about 15 000 km/s (typical kinetic energy of 5 MeV). Therefore, by setting $n - 1 \approx 1.8 \times 10^{-4}$ and $V_0 \approx (15\,000 \pm 2\,000)$ km/s with respect to the surface of the Earth, one finds $δt \approx (60.7 \pm 8.1)$ ns in agreement with the OPERA result.

III. Conclusion

We have investigated the claim for superluminal neutrinos dealing with the OPERA experiment. In view of our analysis, in particular the upper bound we have set to the true velocity of the OPERA neutrinos by using the deformed dispersion relation suggested by the result of the experiment itself, we are led to conclude that the effect unveiled by the OPERA experiment, if confirmed, might not really conflict with Einstein causality. Instead, it can be understood as an apparent faster than light effect that reveals that the impact of the α particles from radioactive rocks of the area of the experiment, as a moving dispersive medium, has not thoroughly been taken into account in computing the velocity of the GPS communication signals through the atmosphere and the magnetosphere. If our suggestion holds, the next generation of positioning systems like Galileo should take it into account.